\documentclass[12pt,a4paper]{article}

\usepackage{graphicx,amsbsy,fancyheadings,supertabular}
\DeclareMathAlphabet{\mathbfit}{OT1}{cmr}{bx}{it}

\newcommand{\bbox}[1]{\mathbfit{#1}}
\newcommand{\bboxs}[1]{\boldsymbol{#1}}

\begin{document}

\title{\large $\enspace$ \\
 \textbf{%\hspace*{-.7cm}
 Radiative Transfer Theory and 
 Diffusion of Light in \hfill $\,$\\
 Nematic Liquid Crystals \hfill $\enspace$}}

\author{\normalsize \hspace*{-.2cm}HOLGER STARK \hfill $\,$\\
%\address{
\normalsize \hspace*{-.2cm}Institut f\"ur Theoretische und Angewandte Physik, 
 Universit\"at Stuttgart, \hfill $\enspace$\\
\normalsize \hspace*{-.2cm}Pfaffenwaldring 57, 70550 Stuttgart, 
Germany \hfill $\,$\\$\enspace$}

%$\enspace$

\date{}

\maketitle

%\begin{abstract}

\hspace*{-.6cm}In nematic liquid crystals light is strongly scattered
from director 
fluctuations. We are interested in the limit where the incoming light wave is 
scattered many times. Then, the light transport can be described by a 
diffusion equation for the energy density of light with diffusion
constants $D_{\|}$ and $D_{\perp}$, respectively, 
parallel and perpendicular to the director. We start from a radiative
transfer theory, connect the diffusion constants to the
dynamic structure factor of  director fluctuations, and shortly
discuss our results.
%their dependence on the Frank elastic constants,
%the dielectric anisotropy, and an applied magnetic field.
%We discuss how $D_{\|}$ and $D_{\perp}$ depend on the Frank elastic constants,
%the dielectric anisotropy, and an applied magnetic field and compare our 
%theoretical predictions with experiment. 
Temporal correlations of the diffusing light probe the dynamics of
director modes on much shorter time scales than single light scattering 
experiments. To account for the decaying temporal correlations, one
has to add an absorption term to the diffusion equation, which we also
link to the dynamic structure factor.
%\end{abstract}

$\enspace$\\

\hspace*{-.6cm}{\em Keywords:\/ } Nematic liquid crystals; light scattering;
radiative transfer theory; diffusion\\
$\enspace$\\

\thispagestyle{empty}

\section*{\normalsize \textbf{1. INTRODUCTION}}

More then one decade ago the discovery of coherent backscattering
or weak localization of light in colloidal suspensions 
[1]
%\cite{Kuga84} 
initiated a tremendous research activity in multiple light scattering
[2].
%\cite{Sheng90}. 
Researchers were attracted by the possibility to achieve the
equivalent to the Anderson localization of electrons in disordered
solids 
[3].
%\cite{Anderson58}. 
So far strong localization of light waves
has not been observed. 

In the theoretical description of coherent backscattering 
the diffusion limit for multiply scattered light is employed 
[4].
%\cite{Golubentsev84}. 
Photons are considered as random walkers with a 
scattering mean free path $l$, measuring the
length they travel between two scattering events, and a diffusion
constant $D = c l^{\ast} / 3$, which involves the transport mean free path
$l^{\ast} = l / \langle  1 - \cos\vartheta_{s} \rangle$. It stands for
the path length beyond which the direction of propagation 
of a photon is randomized. The angular brackets denote an average over all
possible scattering events and $\vartheta_{s}$ is the scattering angle.
In 1987 Wolf and Maret discovered that diffusing light could be used
for spectroscopy 
[5],
%\cite{Maret87}, 
which was later
called Diffusing Wave Spectroscopy (DWS) by Pine {\em et al.\/} 
[6].
%\cite{Pine88}.
This was an important step forward, since so far turbid systems could
not be investigated with conventional dynamic light scattering.
In DWS temporal correlations of the detected intensities decay much
faster than in single scattering since phase shifts of electric field
waves from many scattering events are added up. DWS, therefore,
detects dynamic phenomena at much shorter time scales then normal
dynamic light scattering 
[7].
%\cite{Weitz92}.

Research on diffusing light and DWS has focused on isotropic
media, like colloidal suspensions 
[7,~8],
%\cite{Weitz92,Wu90}, 
emulsions 
[9],
%\cite{Gang94}, 
and foams 
[10].
%\cite{Durian91}. 
Recently, diffusing photons were used for the imaging of objects 
[11,~12]
%\cite{Boas95,Heckmeier96} 
which found its application in medicine 
[13].
%\cite{Yodh95}. 
In magnetic fields a photonic Hall effect was discovered 
[14].
%\cite{Tiggelen95}.
Interest in diffusing light in nematic liquid crystals was
again initiated by the observation of coherent backscattering first in
multidomain samples
[15]
% \cite{Vlasov88}
 and then in a perfectly aligned nematic state
[16].
% \cite{Vithana93}. 
In treating multiply scattered light in the nematic 
phase, one has to deal with the anisotropy in light propagation and a 
different scattering mechanism. Light is not scattered from local objects
like particles in colloidal suspensions but rather from long-range
correlated director fluctuations 
[17,~18].
%\cite{Gennes93,Chandrasekhar92}. 
The theory for diffusing light in nematic liquid crystals was
developed independently by Stark and Lubensky
[19,~20,~21]
% \cite{Stark96a,Stark96b,Stark97}
and Tiggelen, Maynard, and Heiderich 
[22,~23].
%\cite{Tiggelen96,Heiderich97}.
For a review see ref. 
[24].
%\cite{Tiggelen98}.
Its final content can be summarized in an anisotropic diffusion
equation with absorption for the electric field autocorrelation
function $W(\bbox{R},T,t) = \langle \bbox{E}(\bbox{R},T+t/2) \cdot
\bboxs{\varepsilon}_{0} \bbox{E}(\bbox{R},T-t/2) \rangle$:

\begin{equation}
\left[ \frac{\partial}{\partial T} - D_{\|} \nabla^{2}_{\|} - D_{\perp} 
\mathbf{\nabla}^{2}_{\perp} + \mu(t) \right] W(\bbox{R},T,t) = 0
\enspace,
\end{equation}
\\
where $D_{\|}$ and $D_{\perp}$ denote, respectively, the diffusion
constants parallel and perpendicular to the nematic director. The
dynamic absorption coefficient $\mu(t)$, measured in DWS experiments, governs
the temporal decay of the autocorrelation function [$\mu(t=0) =0$].
The mentioned theories link the parameters $D_{\|}$, $D_{\perp}$, and 
$\mu(t)$ to the structure factor of the director fluctuations.
The absorption coefficient $\mu(t)$ appears as an angular average
over all director modes. Careful experiments by Kao {\em et al.\/} 
demonstrated the anisotropic diffusion and the application of
DWS 
[21,~25].
%\cite{Stark97,Kao96}. 
Their results were in excellent agreement with theory.

\chead{LIGHT DIFFUSION IN NEMATICS}

The developed theories used a Green function approach to arrive at
the diffusion equation. In the present article we present
a derivation without the extended formalism of Green functions but
which nevertheless contains the main ideas. To do so we start from
a radiative transfer theory which was first introduced as early as
1905 by Schuster to describe the transport of light through the
atmosphere 
[26].
%\cite{Schuster05}. 
The radiative transfer theory is basically equivalent to
a Boltzmann equation for the energy density of light 
[27,\ 28].
%\cite{Ishimaru78,Hulst80}.
Any coherent superposition of electric field waves are omitted. 
For anisotropic media it needs some refinements which we present in 
section 3. The advantage of the radiative transfer theory is that it
only requires some intuitive understanding. It can, however, be
derived directly from Maxwell's theory. For nematic liquid crystals
this was done first by Romanov and Shalaginov 
[29].
%\cite{Romanov88}.

The article intends to stimulate further research on diffusing light
in liquid crystals.  There exists a wealth of material which scatters
light very strongly, namely porous media filled with nematics 
[30],
%\cite{Bellini96}, 
polymer dispersed liquid crystals 
[31],
%\cite{Drzaic95}, 
focal conic textures in cholesterics
[32],
% \cite{Yang96},
 and the Blue Phase~III. 
Also applications to lyotropic and polymeric liquid crystals and to 
liquid crystalline colloids could be of interest.

Section 2 first summarizes light propagation in a homogeneous nematic phase
emphasizing the differences to isotropic systems. Then we review single
light scattering from director fluctuations and introduce the dynamic
structure factor. Section 3 presents the transport equation for
radiative transfer. An approximation for the diffusion constants
is derived and the formula for the dynamic absorption coefficient is
developed. The main features of both quantities are discussed.

% Radiative Transfer theory: generalization to anisotropic media and
% to temporal correlations \cite{Shalanikov}

% Cite theoretical work and experimental work
% What highlights, further work

% Explain diffusion with absorption: What information is in it?
% range of validity

\section*{\normalsize \textbf{2. LIGHT PROPAGATION AND 
SINGLE SCATTERING IN NEMATIC LIQUID CRYSTALS}}

\chead{H. STARK}

There are two necessary ingredients to treat multiple light scattering
in a turbid medium: the propagation of a plane wave between scattering 
events in an effectively homogeneous system and the single scattering
itself.  We start with the first point.

\subsection*{\normalsize \textbf{A. Light Propagation}}

A homogeneous nematic with the equilibrium Frank director 
$\bbox{n}_{0}$ represents a uniaxial and therefore birefringent material.
It posseses the dielectric tensor 
[17,\ 18]
%\cite{Gennes93,Chandrasekhar92}

\begin{equation}
\label{1.1}
\bboxs{\varepsilon}_{0} = \varepsilon_{\perp} \mathbf{1} + \Delta \varepsilon
\bbox{n}_{0} \otimes \bbox{n}_{0} \enspace ,
\end{equation}
\\
where $\varepsilon_{\|}$ and $\varepsilon_{\perp}$ denote the dielectric 
constants for electric fields, respectively, parallel and perpendicular
to the director, and where $\Delta \varepsilon = \varepsilon_{\|} -
\varepsilon_{\perp}$ stands for the dielectric anisotropy.
There exist two characteristic light modes which we describe by plane
waves for the electric field:
$\bbox{E}(\bbox{r},t) = E^{\alpha} \bbox{e}_{\alpha}(\hat{\bbox{k}}) \,
\exp[i(-\omega t + \bbox{k} \cdot \bbox{r} )]$.
They are characterized by the polarization vector 
$\bbox{e}_{\alpha}(\hat{\bbox{k}})$ and the refractive index 
$n_{\alpha}(\hat{\bbox{k}}) = ck/\omega$ 
[33].
%\cite{Landau60}.
The ordinary light wave
($\alpha = 2$) behaves as in an isotropic medium. It has a
constant index of refraction
$n_{2} = \sqrt{\varepsilon_{\perp}}$ and the polarization vector 
$\bbox{e}_{2}(\hat{\bbox{k}})$ is perpendicular to both $\bbox{n}_{0}$
and $\bbox{k}$.
For the extraordinary mode ($\alpha =1$), however, $n_{1}(\hat{\bbox{k}})$
depends on the angle $\vartheta$ between $\bbox{k}$ and $\bbox{n}_{0}$:

\begin{equation}
\label{1.2}
\frac{1}{n_{1}^{2}(\hat{\bbox{k}})} = 
\frac{\sin^{2}\vartheta}{\varepsilon_{\|}} +
\frac{\cos^{2}\vartheta}{\varepsilon_{\perp}} \enspace.
\end{equation}
\\
The vector $\bbox{e}_{1}(\hat{\bbox{k}})$ lies in the plane defined by
$\bbox{n}_{0}$ and $\bbox{k}$ but is not perpendicular to the wave
vector $\bbox{k}$. It makes sense to introduce the polarization vector
$\bbox{d}^{\alpha}(\hat{\bbox{k}}) = \bbox{\varepsilon}_{0}
\bbox{e}_{\alpha}(\hat{\bbox{k}})$ for the electric displacement field.
Then, the biorthogonality relation 
$\bbox{d}^{\alpha} \cdot \bbox{e}_{\beta} =
\delta^{\alpha}_{\beta}$ holds
[20,\ 34].
% \cite{Stark96b,Lax71}. 
The normalization
of the polarization vectors is chosen such that the energy density 
$\bbox{E}\cdot \bboxs{\varepsilon}_{0} \bbox{E} /4\pi$ of a light mode,
averaged over on time period, becomes $W^{\alpha} = |E^{\alpha}|^{2} / 8\pi$.

Besides the phase velocity $\bbox{v}_{p \alpha} = c \hat{\bbox{k}}/
n_{\alpha}$ there exists the group velocity 
$\bbox{v}_{g\alpha} = \bboxs{\nabla }_{\bbox{k}} \omega_{\alpha}(\bbox{k})$,
where $\omega_{\alpha}(\bbox{k}) = c k / n_{\alpha}(\hat{\bbox{k}})$
stands for the dispersion relation. Electromagnetic energy is
transported along the direction of $\bbox{v}_{g\alpha}$. For ordinary
light $\bbox{v}_{g2} = \bbox{v}_{p2}$. For extraordinary light the group 
velocity

\begin{equation}
\label{1.3}
\bbox{v}_{g1} = c n_{1}(\hat{\bbox{k}}) \, 
\left( \frac{\cos \vartheta}{\varepsilon_{\perp}} \bbox{n}_{0} +
       \frac{\sin \vartheta}{\varepsilon_{\|}} \hat{\bbox{u}}_{1} \right)
\end{equation}
\\
differs in direction and magnitude from the phase velocity. We
introduced the unit vector $\hat{\bbox{u}}_{1}$ perpendicular to both
$\bbox{n}_{0}$ and $\bbox{e}_{2}(\hat{\bbox{k}})$.
%lies in the plane defined by both $\bbox{n}_{0}$ and $\bbox{k}$. 
In a medium without absorption the energy transport is described by
the Poynting vector
$c(\bbox{E} \times \bbox{H}) / 4\pi$. It can be rewritten so that for
each light mode it equals the energy density times the group velocity:
$\bbox{S}^{\alpha} = W^{\alpha} \bbox{v}_{g\alpha}$ 
[33].
%\cite{Landau60}.

\subsection*{\normalsize \textbf{B. Single Scattering}}

\chead{LIGHT DIFFUSION IN NEMATICS}

Single light scattering from thermally activated director modes has been well
understood for a long time 
[17,\ 18].
%\cite{Gennes93,Chandrasekhar92}. 
The fluctuating part of the director,
$\delta \bbox{n}(\bbox{r},t)$, induces fluctuations in the dielectric
tensor, $\delta \bboxs{\varepsilon} = \Delta \varepsilon 
[\delta \bbox{n} \otimes \bbox{n}_{0} +
 \bbox{n}_{0} \otimes \delta \bbox{n}]$, which scatter light. 
The typical scattering experiment involves incoming light with wave vector 
$ \bbox{k}^{\alpha} = \omega n_{\alpha} \hat{\bbox{k}} /c$
and with polarization $\bbox{e}_{\alpha}(\hat{\bbox{k}})$ which is
partially scattered into light with wave vector
$\bbox{q}^{\beta} = \omega n_{\beta} \hat{\bbox{q}} /c$ and
polarization $\bbox{e}_{\beta}(\hat{\bbox{q}})$. In the
weak-scattering approximation, the temporal
autocorrelation function of the scattered electric field is
proportional to the dynamic structure factor

\begin{equation}
\label{1.4}
B_{\alpha \beta}(\hat{\bbox{k}}, \hat{\bbox{q}}, t) = 
\frac{\omega^{4}}{c^{4}} 
\langle \, \delta \varepsilon_{\alpha \beta}(\bbox{q}_{s},t) \,
\delta \varepsilon_{\alpha \beta}^{\ast}(\bbox{q}_{s},0) \, \rangle 
\enspace ,
\end{equation}
\\
where $\bbox{q}_{s} = \bbox{q}^{\beta} - \bbox{k}^{\alpha}$ denotes
the scattering vector and $\delta \varepsilon_{\alpha \beta}(\bbox{q}_{s},t) =
$\linebreak$\bbox{e}_{\beta}(\hat{\bbox{q}}) \cdot 
\delta \bboxs{\varepsilon}(\bbox{q}_{s},t) \bbox{e}_{\alpha}(\hat{\bbox{k}})
$ stands for the projection of $\delta \bboxs{\varepsilon}$ on the
polarization vectors. In its final form we obtain for the structure factor:

\begin{equation}
\label{1.5}
B_{\alpha \beta}(\hat{\bbox{k}}, \hat{\bbox{q}}, t) = (\Delta \varepsilon)^{2}
\frac{\omega^{4}}{c^{4}} \sum_{\delta =1}^{2}
N(\alpha,\beta,\delta) \frac{k_{\mathrm{B}} T}{K_{\delta}(\bbox{q}_{s})} \,
\exp\left[-\frac{K_{\delta}(\bbox{q}_{s})}{\eta_{\delta}(\bbox{q}_{s})}
 t \right] \enspace.
\end{equation}
\\
The term $k_{\mathrm{B}} T / K_{\delta}(\bbox{q}_{s})$ represents the
thermally activated director modes. It involves the
elastic coefficient $K_{\delta}(\bbox{q}_{s}) = K_{\delta} q^{2}_{\perp} +
K_{3} q^{2}_{\|} + \Delta \chi H^{2}$ with Frank constants $K_{i}$,
magnetic field $H$, magnetic anisotropy $\Delta \chi$, and the components
of $\bbox{q}_{s} = (\bbox{q}_{\perp} , q_{\|})$. The exponential factor
reflects the diffusive nature of the director modes. The relaxation
frequency depends on the viscosity $\eta_{\delta}(\bbox{q}_{s}) = \gamma
- \mu(\bbox{q}_{s})$, where the rotational viscosity $\gamma$ plays the
important role. Finally, $N(\alpha,\beta,\delta)$ stands for a 
geometry factor which forbids ordinary-to-ordinary scattering and forward
scattering along the director. From the structure factor we derive
the scattering mean free path $l_{\alpha}(\hat{\bbox{k}})$
[20,\ 22,\ 35]:
%\cite{Langevin75,Stark96b,Tiggelen96}:

\chead{H. STARK}

\begin{equation}
\label{1.6}
\frac{1}{l_{\alpha}(\hat{\bbox{k}})} = n_{\alpha}(\hat{\bbox{k}})
\sum_{\beta} \int \frac{d\Omega_{\bbox{q}}}{(4\pi)^{2}} 
B_{\alpha \beta}(\hat{\bbox{k}}, \hat{\bbox{q}}, 0) 
n_{\beta}^{3}(\hat{\bbox{q}}) \enspace,
\end{equation}
\\
which depends on the polarization $\alpha$ and the direction $\hat{\bbox{k}}$
of light. In the photon picture it gives the average path length a
photon travels between two scattering events. For light intensity
it gives the distance after which the initial intensity $I_{0}$ has
decayed to $I_{0}/e$. However, multiple scattering events are totally
neglected in the last interpretation. For a careful interpretation
of the scattering mean free path in connection with $\hat{\bbox{k}}$ and
$\bbox{v}_{g\alpha}$ see ref.\
[21].
% \cite{Stark97}. 
The mean free path is discussed in refs.\
[24,\ 35].
% \cite{Langevin75,Tiggelen98}.
We only note, that $l_{1}(\hat{\bbox{k}})$ goes to zero when $H
\rightarrow 0$ since the structure factor for
extraordinary-to-extraordinary scattering diverges for $\bbox{q}_{s}
\rightarrow \mathbf{0}$.

Finally, we make some comments about the symmetry of the structure
factor. In isotropic systems $B(\hat{\bbox{k}} \cdot \hat{\bbox{q}})$
only depends on the scattering angle $\vartheta_{s}$
via $\hat{\bbox{k}} \cdot \hat{\bbox{q}} = \cos
\vartheta_{s}$. Therefore, if expanded into spherical harmonics
$Y_{l'm'}(\hat{\bbox{k}})$ and $Y_{lm}(\hat{\bbox{q}})$, it is 
fully diagonal: $\langle l'm' | B(\hat{\bbox{k}} \cdot \hat{\bbox{q}}) | lm
\rangle \propto \delta_{ll'} \delta_{mm'}$. The structure factor for
the director modes just posseses the rotational symmetry around the
equilibrium director $\bbox{n}_{0}$. It can be written as a
function of the relative azimuthal angle
$\varphi = \varphi_{\bbox{q}} - \varphi_{\bbox{k}}$ between
$\hat{\bbox{q}}$ and $\hat{\bbox{k}}$. Using the additional symmetry
$B_{\alpha \beta}(\hat{\bbox{k}},\hat{\bbox{q}},0) = 
B_{\alpha \beta}(-\hat{\bbox{k}},-\hat{\bbox{q}},0)$ the following
expansion holds:

\begin{equation}
\label{1.7}
B_{\alpha \beta}(\hat{\bbox{k}},\hat{\bbox{q}},0) =
\sum_{m \ge 0} B_{\alpha \beta}^{m}(\vartheta_{\bbox{k}},\vartheta_{\bbox{q}})
\cos[m(\varphi_{\bbox{q}} - \varphi_{\bbox{k}})] \enspace,
\end{equation}
\\
where $\vartheta_{\bbox{k}}$ and $\vartheta_{\bbox{q}}$ are polar angles
of $\hat{\bbox{k}}$ and $\hat{\bbox{q}}$
with respect to $\bbox{n}_{0}$. It expresses the fact that 
$\langle l'm'| B_{\alpha \beta}(\hat{\bbox{k}},\hat{\bbox{q}},0)| lm
\rangle \propto \delta_{mm'}$ is diagonal in the index $m$ but
not in $l$.

\section*{\normalsize 
\textbf{3. RADIATIVE TRANSFER THEORY AND DIFFUSION APPROXIMATION}}
%\textbf{{\large 3.} RADIATIVE TRANSFER THEORY AND DIFFUSION APPROXIMATION}}

In the following we deal with the temporal auto correlation function of the
electric field

\begin{equation}
\label{2.1}
%W_{\hat{\bbox{k}}}^{\alpha}(\bbox{R},T,t) =
n^{3}_{\alpha}(\hat{\bbox{k}})
W_{\hat{\bbox{k}}}^{\alpha}(\bbox{R},T,t)
= \langle \, E^{\alpha}_{\hat{\bbox{k}}}(\bbox{R},T-t/2) \, 
E^{\alpha\ast}_{\hat{\bbox{k}}}(\bbox{R},T+t/2) \, \rangle \enspace.
\end{equation}
\\
For $t=0$, it stands for the energy density of a light wave at time $T$
and space point $\bbox{R}$ travelling into direction $\hat{\bbox{k}}$ with
polarization $\alpha$. The light frequency $\omega$ is omitted.
We pulled out a factor $n^{3}_{\alpha}(\hat{\bbox{k}})$.
It is proportional to the number $N_{\alpha}(\omega,\hat{\bbox{k}}) d \omega 
d\Omega_{\bbox{k}}$ of photon states for a given polarization $\alpha$,
frequency $\omega$, and direction $\hat{\bbox{k}}$, since 
$N_{\alpha}(\omega,\hat{\bbox{k}}) d \omega d\Omega_{\bbox{k}}
\propto k^{2} d k d\Omega_{\bbox{k}} = 
n^{3}_{\alpha}(\hat{\bbox{k}})\omega^{2} d\omega d\Omega_{\bbox{k}} / c^{3}$.

\chead{LIGHT DIFFUSION IN NEMATICS}

%in a frequency intervall $[\omega , \omega + d \omega]$, since 
%$N(\omega,\hat{\bbox{k}})$ is proportional to the volume element 
%$ k^{2} dk d\Omega = n^{3}_{\alpha}(\hat{\bbox{k}}) \omega^{2} d\omega
%d\Omega / c^{3}$ in $\bbox{k}$ space.

The equation of radiative transfer theory formally corresponds to a
Boltzmann equation balancing all the changes in 
$W_{\hat{\bbox{k}}}^{\alpha}(\bbox{R},T,t)$:

\begin{eqnarray}
\label{2.2}
\lefteqn{\left(\frac{\partial}{\partial T} + \bbox{v}_{g\alpha} \cdot
\bboxs{\nabla} + \frac{1}{l_{\alpha}}\frac{c}{n_{\alpha}}\right)
 W_{\hat{\bbox{k}}}^{\alpha}(\bbox{R},T,t) \qquad \qquad} \nonumber \\
 & & = c \sum_{\beta} \int \frac{d \Omega_{\bbox{q}}}{(4\pi)^{2}}
B_{\alpha\beta}(\hat{\bbox{k}},\hat{\bbox{q}},t) n^{3}_{\beta}(\hat{\bbox{q}})
W_{\hat{\bbox{q}}}^{\beta}(\bbox{R},T,t) + 
S^{\alpha}_{\hat{\bbox{k}}}(\bbox{R},T) \enspace.
\end{eqnarray}
\\
The first term gives temporal variations of
$W_{\hat{\bbox{k}}}^{\alpha}(\bbox{R},T,t)$ due to e.g. time dependent
light sources. The second term involves the divergence of the Poynting
vector $\bbox{v}_{g\alpha} W_{\hat{\bbox{k}}}^{\alpha}(\bbox{R},T,t)$.
The correlation function $W_{\hat{\bbox{k}}}^{\alpha}(\bbox{R},T,t)$
changes when there is a net flow of energy ($t=0$) or correlation 
($t \ne 0$) out of the volume element around $\bbox{R}$.
The third and fourth terms describe losses and
gains due to scattering. The scattering mean free path 
$l_{\alpha}(\hat{\bbox{k}})$ has been already introduced in Eq.~(\ref{1.6}).
Finally, $S^{\alpha}_{\hat{\bbox{k}}}(\bbox{R},T)$ indicates a source term. 
The transport equation~(\ref{2.2}) is intuitively understandable.
Compared to standard textbooks on multiple light scattering 
[27,\ 28]
%\cite{Ishimaru78,Hulst80} 
it contains two generalizations. First, it is
valid for general anisotropic random media. Secondly, for $t \ne
0$, it describes the transport of electric field correlations 
[11,\ 36].
%\cite{Boas95,Ackerson92}.
The transport equation can be derived from first principles,
i.~e. starting from Maxwell's theory. It follows in a straightforward
way from the Bethe-Salpeter equation for the averaged "two-particle"
Green function 
[19,\ 20,\ 22,\ 29].
%\cite{Stark96a,Stark96b,Tiggelen96,Romanov88}. 
Its validity is restricted to length and time scales much longer than the
wavelength and the time period of light.

\subsection*{\normalsize \textbf{A. Diffusion Approximation}}

To derive the diffusion approximation from Eq.~(\ref{2.2}) we set
$t=0$ and neglect any source terms. We first study the equilibrium
solution of Eq.~(\ref{2.2}), where
$W_{\hat{\bbox{k}}}^{\alpha}(\bbox{R},T)$ equals a constant $W_{0}$.
As a result formula (\ref{1.6}) for the scattering mean free path
$l_{\alpha}(\hat{\bbox{k}})$ is reproduced. The energy density
$ n^{3}_{\alpha}(\hat{\bbox{k}}) W_{0}$ still depends
on the direction $\hat{\bbox{k}}$ and polarization $\alpha$ of light,
due to the equipartition of the light energy on all available photon states.
So far, this still needs an experimental confirmation.
The diffusion approximation follows when the $\hat{\bbox{k}}$
dependence of $W_{\hat{\bbox{k}}}^{\alpha}(\bbox{R},T)$ deviates only
slightly from the equilibrium angular distribution 
[27,\ 28].
%\cite{Ishimaru78,Hulst80}.
Let us therefore look at an angular expansion of 
$W_{\hat{\bbox{k}}}^{\alpha}(\bbox{R},T)$:

\chead{H. STARK}

\begin{eqnarray}
W_{\hat{\bbox{k}}}^{\alpha}(\bbox{R},T) & = & \frac{1}{8\pi} W_{0}(\bbox{R},T)
+ \frac{3}{4\pi} \frac{1}{c \overline{n^{3}_{\alpha}}} \,
n_{\alpha} \hat{\bbox{k}} \cdot \bbox{J}^{\alpha}(\bbox{R},T) \nonumber\\
\label{2.3}
 & & 
%\qquad \qquad \qquad 
+ \sum_{l>1,m} W^{\alpha}_{lm}(\bbox{R},T) 
\widetilde{Y}_{lm}^{\alpha}(\hat{\bbox{k}}) \enspace,
\end{eqnarray}
\\
where we introduced the total energy density

\begin{equation}
\label{2.4}
\overline{n^{3}} W_{0}(\bbox{R},T) = \sum_{\alpha} \int d\Omega_{\bbox{k}} 
n^{3}_{\alpha}(\hat{\bbox{k}}) W^{\alpha}_{\hat{\bbox{k}}}(\bbox{R},T)
\end{equation}
\\
as $W_{0}$ times an angular average over the cubes of both
refractive indices:

\begin{equation}
\label{2.5}
\overline{n^{3}} = (\overline{n_{1}^{3}} + \overline{n_{2}^{3}}) / 2 = 
\int d\Omega_{\bbox{k}} 
[n^{3}_{1}(\hat{\bbox{k}}) + n^{3}_{2}(\hat{\bbox{k}})] / 8\pi =
( \varepsilon_{\perp}^{1/2} \varepsilon_{\|} +
\varepsilon_{\perp}^{3/2} )/2 \enspace.
\end{equation}
\\
Our goal is to establish a diffusion equation for $W_{0}(\bbox{R},T)$.
We also defined the total energy density current 
$\bbox{J}^{\alpha}(\bbox{R},T)$ of light with polarization~$\alpha$,

\begin{equation}
\label{2.6}
\bbox{J}^{\alpha}(\bbox{R},T) = \int d\Omega_{\bbox{k}} n^{3}_{\alpha}
\bbox{v}_{g\alpha} W^{\alpha}_{\hat{\bbox{k}}}(\bbox{R},T) \enspace.
\end{equation}
\\
A few comments are necessary to understand Eq.~(\ref{2.3}).
In isotropic systems, the second term on the right-hand side just
corresponds to an expansion into the components of 
$\hat{\bbox{k}} = (\sin \vartheta \cos \varphi, \sin \vartheta \sin \varphi,
\cos \vartheta)$, which basically stand for $l=1$ spherical harmonics.
Here, for uniaxial systems, it is useful to choose basis functions
with $n_{\alpha}^{3}(\hat{\bbox{k}})$ as a weight function:

\begin{equation}
\label{2.7}
\int d\cos\vartheta d\varphi \, n_{\alpha}^{3}(\hat{\bbox{k}})
\widetilde{Y}_{lm}^{\alpha}(\hat{\bbox{k}})
\widetilde{Y}_{l'm'}^{\alpha}(\hat{\bbox{k}}) = \overline{n^{3}_{\alpha}}
\delta_{ll'} \delta_{mm'} \enspace.
\end{equation}
\\
We will see below that this choice establishes an approximation scheme
for the diffusion constants of light.
For ordinary waves we still obtain the conventional spherical harmonics.
For extraordinary waves we introduce a new coordinate
$C = n_{1}(\hat{\bbox{k}}) \cos\vartheta / n_{2}$,
which is equivalent to $\bbox{n}_{0} \cdot \hat{\bbox{k}} = 
\cos\vartheta$ since it also ranges from $-1$ to $1$. 
With this choice the weight function becomes a constant:

\chead{LIGHT DIFFUSION IN NEMATICS}

\begin{equation}
\label{2.8}
\int d \cos \vartheta n^{3}_{1}(\hat{\bbox{k}}) \ldots =
\overline{n^{3}_{1}}  \int dC \ldots
\end{equation}
\\
Hence, the basis functions $\widetilde{Y}_{lm}^{1}(\hat{\bbox{k}})$
for $\alpha =1$
simply follow from spherical harmonics when $\cos \vartheta$ is replaced by 
$C$. With the abbreviation $C=\cos \vartheta$ for ordinary light, the 
generalized spherical harmonics are the same for $\alpha=1$ and 2.
For $l=1$ in real representation, they read:

\begin{equation}
\widetilde{Y}^{\alpha}_{10}(\hat{\bbox{k}}) = \sqrt{\frac{3}{4\pi}} C \quad
, \quad \frac{\widetilde{Y}^{\alpha}_{11}(\hat{\bbox{k}}) \pm
              \widetilde{Y}^{\alpha}_{1-1}(\hat{\bbox{k}}) }{2} =
\sqrt{\frac{3}{4\pi}} \sqrt{1-C^{2}} 
\left\{ \begin{array}{c} \cos\varphi \\ \sin{\varphi} \end{array}  \right.
\end{equation}
\\
In expansion~(\ref{2.3}) we have already used them explicitly in the 
second term. We also note that $n_{1}(\hat{\bbox{k}}) \sin \vartheta / n_{2} =
\sqrt{\varepsilon_{\|\rule[-1.5mm]{0mm}{2mm}} / 
\varepsilon_{\perp}} \sqrt{1-C^{2}}$.
%$n_{1}(\hat{\bbox{k}}) \sin \vartheta / n_{2} =
%( \varepsilon_{\|} / \varepsilon_{\perp})^{1/2} (1-C^{2})^{1/2}$

We are ready to extract the diffusion approximation from the transport
equation~(\ref{2.2}). Multiplying Eq.~(\ref{2.2}) by
$n_{\alpha}^{3}$ and summing over all directions of $\hat{\bbox{k}}$
and the two polarizations leads to the continuity equation for the 
energy density

\begin{equation}
\label{2.9}
\frac{\partial }{\partial T} \overline{n^{3}} W_{0} + \bboxs{\nabla}
  \cdot \bbox{J} = 0 \enspace.
\end{equation}
\\
The vector $\bbox{J} = \bbox{J}^{1} + \bbox{J}^{2}$ denotes the total energy 
density current.
A second equation, Fick's law, relates $\bbox{J}$ to the gradient 
of the energy density:

\begin{equation}
\label{2.10}
\bbox{J} = - \bbox{D} \bboxs{\nabla} \overline{n^{3}}W_{0}
\enspace.
\end{equation}
\\
It will be derived below. We introduced the diffusion tensor

\begin{equation}
\label{2.11}
\bbox{D} = D_{\perp} \mathbf{1} + (D_{\|} - D_{\perp}) \bbox{n}_{0}
\otimes \bbox{n}_{0}
\end{equation}
\\
with its two independent light diffusion constants $D_{\|}$ and 
$D_{\perp}$, respectively, parallel and perpendicular to the director
$\bbox{n}_{0}$. Eliminating the current $\bbox{J}$
finally gives the diffusion equation for $W_{0}(\bbox{R},T)$,

\begin{equation}
\label{2.12}
\left( \frac{\partial}{\partial T} - D_{\|} \nabla_{\|}^{2} - 
D_{\perp} \bboxs{\nabla}_{\perp}^{2} \right) W_{0}(\bbox{R},T) = 0
\enspace,
\end{equation}
\\
where $\bboxs{\nabla} = (\bboxs{\nabla}_{\perp} , 
\nabla_{\|\rule[-1.5mm]{0mm}{2mm}} )$.
%where $\nabla_{\|}$ and $\bboxs{\nabla}_{\perp}$, respectively, denote
%the components of $\bboxs{\nabla}$ parallel and perpendicular to the
%director $\bbox{n}_{0}$.

\chead{H. STARK}

To obtain Fick's law\ (\ref{2.10}), we first look at components parallel to
the director $\bbox{n}_{0}$. We project the transport equation~(\ref{2.2})
on $ \widetilde{Y}^{\alpha}_{10}(\hat{\bbox{k}}) \propto C_{\bbox{k}}$
%where $\vartheta_{\bbox{k}}$ indicates the polar angle of $\hat{\bbox{k}}$
%with respect to $\bbox{n}_{0}$, 
and arrive at a set of equations which couple the energy density $W_{0}$ and 
the components $J^{1}_{\|}$ and $J^{2}_{\|}$:

\begin{equation}
\label{2.13}
\frac{(4\pi)^{3}}{18} \frac{c}{n_{2}^{2}} \nabla_{\|} W_{0}
\left( \begin{array}{c} 1 \\ 1 \end{array} \right) +
\left( \begin{array}{cc}
         {\cal B}_{11}^{\|} & {\cal B}_{12}^{\|} \\
         {\cal B}_{12}^{\|} & {\cal B}_{22}^{\|}
       \end{array} \right) \,
\left( \begin{array}{c} J^{1}_{\|} \\ J^{2}_{\|} \end{array} \right) +
\dots = \mathbf{0} \enspace.
\end{equation}
\\
% The matrix elements ${\cal B}_{\alpha \beta}^{\|}$ of 
%$B_{\alpha \beta}(\hat{\bbox{k}},\hat{\bbox{q}})$ stand for
The quantitites ${\cal B}_{\alpha \beta}^{\|}$ are extended matrix elements
of $B_{\alpha \beta}(\hat{\bbox{k}},\hat{\bbox{q}})$:

\begin{eqnarray}
{\cal B}_{11}^{\|} & = & \int_{\hat{\bbox{k}}} \int_{\hat{\bbox{q}}}
%D(\hat{\bbox{k}},\hat{\bbox{q}})
%dC_{\bbox{k}} d \varphi_{\bbox{k}} dC_{\bbox{q}} d \varphi_{\bbox{q}}
\, [(C_{\bbox{k}}^{2} - C_{\bbox{k}}C_{\bbox{q}}) 
B_{11}(\hat{\bbox{k}},\hat{\bbox{q}}) + 
C_{\bbox{k}}^{2} \frac{\varepsilon_{\perp}}{\varepsilon_{\|}} 
B_{12}(\hat{\bbox{k}},\hat{\bbox{q}})] \nonumber \\
\label{2.14}
{\cal B}_{22}^{\|} & = & \int_{\hat{\bbox{k}}} \int_{\hat{\bbox{q}}} 
%D(\hat{\bbox{k}},\hat{\bbox{q}})
%dC_{\bbox{k}} d \varphi_{\bbox{k}} dC_{\bbox{q}} d \varphi_{\bbox{q}}
\, [(C_{\bbox{k}}^{2} - C_{\bbox{k}}C_{\bbox{q}}) 
B_{22}(\hat{\bbox{k}},\hat{\bbox{q}}) + 
C_{\bbox{k}}^{2} \frac{\varepsilon_{\perp}}{\varepsilon_{\|}} 
B_{21}(\hat{\bbox{k}},\hat{\bbox{q}})] \\
{\cal B}_{12}^{\|} & = & - \int_{\hat{\bbox{k}}} \int_{\hat{\bbox{q}}}
%D(\hat{\bbox{k}},\hat{\bbox{q}})
%dC_{\bbox{k}} d \varphi_{\bbox{k}} dC_{\bbox{q}} d \varphi_{\bbox{q}} 
C_{\bbox{k}} C_{\bbox{q}} 
B_{12}(\hat{\bbox{k}},\hat{\bbox{q}}) \nonumber
\end{eqnarray}
\\
where we used the abbreviation 
$\int_{\hat{\bbox{k}}} = \int dC_{\bbox{k}} d \varphi_{\bbox{k}} $.
In Eq.~(\ref{2.13}) we neglected terms proportional to 
$\partial J^{\alpha}_{\|} / \partial T$. We
also assumed that further contributions containing $W^{\alpha}_{l0}$
with $l>1$ are small and discuss this approximation below.
Solving for the two currents $J_{\|}^{1}$ and $J_{\|}^{2}$ leads to
the parallel component of Eq.~(\ref{2.10}):

\begin{equation}
\label{2.15}
J_{\|} = J_{\|}^{1} + J_{\|}^{2} = - D_{\|} \nabla_{\|}
\overline{n^{3}} W_{0}
\end{equation}
\\
with the diffusion constant

\begin{equation}
\label{2.16}
D_{\|} = \frac{(4\pi)^{3}}{18} \, \frac{c}{n_{2}^{2} \overline{n^{3}}} \,
\frac{{\cal B}_{11}^{\|} + {\cal B}_{22}^{\|} -2 {\cal B}_{12}^{\|}}{
{\cal B}_{11}^{\|} {\cal B}_{22}^{\|} - ({\cal B}_{12}^{\|})^{2} } \enspace.
\end{equation}
\\

\chead{LIGHT DIFFUSION IN NEMATICS}

The perpendicular component of Eq.~(\ref{2.10}) follows in an analogous
way after projecting the transport equation~(\ref{2.2}) on 
$[\widetilde{Y}_{11}^{\alpha}(\hat{\bbox{k}}) + 
\widetilde{Y}_{1-1}^{\alpha}(\hat{\bbox{k}}) ] \propto$\linebreak
$\sqrt{1-C_{\bbox{k}}^{2}} \cos \varphi_{\bbox{k}}$:

\begin{equation}
\label{2.17}
J_{\perp} = J_{\perp}^{1} + J_{\perp}^{2} = - D_{\perp} \bboxs{\nabla}_{\perp}
\overline{n^{3}} W_{0}
\end{equation}
with

\begin{equation}
\label{2.18}
D_{\perp} = \frac{(4\pi)^{3}}{18} \, \frac{c}{n_{2}^{2} \overline{n^{3}}} \,
\frac{{\cal B}_{11}^{\perp} + {\cal B}_{22}^{\perp}
  \varepsilon_{\perp}/ \varepsilon_{\|} - 2 {\cal B}_{12}^{\perp}
  \sqrt{\varepsilon_{\perp}/\varepsilon_{\|}}}{
{\cal B}_{11}^{\perp} {\cal B}_{22}^{\perp} - ({\cal B}_{12}^{\perp})^{2} }
\enspace.
\end{equation}
\\
The matrix elements ${\cal B}_{\alpha \beta}^{\perp}$ are defined as
in Eqs.~(\ref{2.14}) but with $C$ replaced by $\sqrt{1-C^{2}} \cos \varphi$.

The expressions~(\ref{2.16}) and (\ref{2.18}) for the diffusion
constants are approximate formulas since in Eq.~(\ref{2.13}) we have
neglected terms containing $W^{\alpha}_{l0}$ with $l>1$. In 
ref.\
[20]
%~\cite{Stark96b}
we showed that corrections of $D_{\|}$ and $D_{\perp}$ resulting
from l=3 spherical harmonics are essentially 1\% or smaller.
One could ask for the reason.
The additional coefficients $W^{\alpha}_{l0}$ in Eq.~(\ref{2.13}) with
the non-diagonal matrix elements $\langle 10 | B_{\alpha
  \beta}(\hat{\bbox{k}},\hat{\bbox{q}},0)|l0 \rangle$ as prefactors have
to be determined by projecting the transport equation on higher generalized
spherical harmonics ($l>1$). 
Because of the choice of our basis functions the 
resulting equations only couple $W^{\alpha}_{l0}$ to $J_{\|}^{\beta}$.
The coefficient $W_{0}$ does not appear. That means 
$W^{\alpha}_{l0}$ ($l > 1$)
depends in a complicated way but directly on $J_{\|}^{\beta}$. As a result
the matrix elements ${\cal B}^{\|}_{\alpha \beta}$ are renormalized.
However, the renormalization is small if the
non-diagonal elements $\langle 10 | B_{\alpha
  \beta}(\hat{\bbox{k}},\hat{\bbox{q}},0)|l0\rangle$ are small.
For symmetry reasons it is clear that $D_{\|}$ and $D_{\perp}$
involve, respectively, $m=0$ or $|m| = 1$ spherical harmonics only.
In isotropic systems higher spherical harmonics do not contribute at
all, since $\langle 10 | B(\hat{\bbox{k}} \cdot
\hat{\bbox{q}},0)|l0\rangle \propto \delta_{l1}$, and 
$D \propto \langle 1 - \cos \vartheta_{s} \rangle^{-1}$ follows.

An extensive discussion of the diffusion constants and their
dependence on Frank constants $K_{i}$, dielectric anisotropy 
$\Delta \varepsilon$, and applied magnetic field $H$ is given in 
refs.\
[20,\ 21]
% \cite{Stark96b,Stark97}
 and by Tiggelen {\em et al.\em} 
[22].
%\cite{Tiggelen96}. 
We just summarize the
important results. For a typical material, 5CB, we find 
$D_{\|} = 1.43 \times 10^{9} \mathrm{cm}^{2}/\mathrm{s}$ and
$D_{\perp} = 0.98 \times 10^{9} \mathrm{cm}^{2}/\mathrm{s}$ with a ratio
of $D_{\|}/D_{\perp} = 1.45$ in excellent agreement with experiment
and numerical simulations 
[21,\ 25].
%\cite{Stark97,Kao96}. 
The values were calculated in the
limit of $H \rightarrow 0$, which demonstrates that the divergence
of the structure factor for $\bbox{q}_{s} \rightarrow \mathbf{0}$
does not affect the diffusion constants.
If one introduces transport mean free paths via $ l^{\ast}_{\|/\perp}
= 3 n_{2} D_{\|/\perp} / c $, one arrives at
$l_{\|} = 2.2 \mathrm{mm}$ and $l_{\perp} = 1.5 \mathrm{mm}$ for 5CB. We
stress that it is not obvious how to define transport mean free paths
in anisotropic turbid media. The calculated values give an orientation
only. In the case of $\Delta \varepsilon = 0$ and the one-constant
approximation $K_{i} = K$ there is still a remaining anisotropy of
$D_{\|} / D_{\perp} = 1.06$ due to the inherent anisotropy in the
nematic structure factor. For $\Delta \varepsilon < 0$ there exists a
point with $D_{\|} = D_{\perp}$ beyond which the ratio 
$D_{\|} / D_{\perp}$ is smaller than 1. This behavior should be observable in
discotic nematics.

\chead{H. STARK}

\subsection*{\normalsize \textbf{B. Dynamic Absorption}}

Now, we set $t \ne 0$ and look at the transport of electric field
correlations. In doing so, we restrict ourselves to times $t$ much
smaller than typical director relaxation times 
$\tau = \gamma / (K q_{s}^{2})$. 
For wave numbers $q_{s}$ of light they cover the range 
$\tau = 10 - 100 \,\mu \mathrm{s}$. We rewrite the dynamical structure
factor,

\begin{equation}
\label{2.19}
B_{\alpha \beta}(\hat{\bbox{k}},\hat{\bbox{q}},t) =
B_{\alpha \beta}(\hat{\bbox{k}},\hat{\bbox{q}},0) + 
[B_{\alpha \beta}(\hat{\bbox{k}},\hat{\bbox{q}},t) -
 B_{\alpha \beta}(\hat{\bbox{k}},\hat{\bbox{q}},0) ] \enspace,
\end{equation}
\\
where the second part on the right-hand side assumes the form 

\begin{equation}
\label{2.20}
B_{\alpha \beta}(\hat{\bbox{k}},\hat{\bbox{q}},t) -
B_{\alpha \beta}(\hat{\bbox{k}},\hat{\bbox{q}},0) = -
(\Delta \varepsilon)^{2} k_{\mathrm{B}} T \frac{\omega^{4}}{c^{4}} 
\sum_{\delta =1}^{2} \frac{N(\alpha,\beta,\delta)}{\eta_{\delta}
(\bbox{q}_{s})} \, t
\end{equation}
\\
after an expansion of the exponential factor in 
$B_{\alpha \beta}(\hat{\bbox{k}},\hat{\bbox{q}},t)$.
All the coefficients in an angular expansion of 
$W_{\hat{\bbox{k}}}^{\alpha}(\bbox{R},T,t)$ now carry the
relative time $t$ as a further argument. The important quantity
is the total autocorrelation function 

\begin{equation}
\label{2.21}
\overline{n^{3}} W_{0}(\bbox{R},T,t) = \sum_{\alpha} \int d\Omega_{\bbox{k}} 
n^{3}_{\alpha}(\hat{\bbox{k}}) W^{\alpha}_{\hat{\bbox{k}}}(\bbox{R},T,t)
\enspace.
\end{equation}
\\
In repeating the derivation of the continuity equation~(\ref{2.9}) we
have to add a dynamic absorption term 
$\mu(t) \overline{n^{3}} W_{0}(\bbox{R},T,t)$:

\begin{equation}
\label{2.22}
\left[ \frac{\partial }{\partial T} + \mu(t) \right]    
\overline{n^{3}} W_{0}(\bbox{R},T,t) + \bboxs{\nabla}
  \cdot \bbox{J}(\bbox{R},T,t) = 0 \enspace,
\end{equation}
\\
which means that temporal correlations are not conserved quantities
because they decay to zero. The dynamic absorption coefficient 
$\mu(t)$ follows generally from an angular average over all dynamic
modes of a system:

\chead{LIGHT DIFFUSION IN NEMATICS}

\begin{equation}
\mu(t) = \frac{c}{8 \pi \overline{n^{3}}} \sum_{\alpha \beta} \int \! \int 
\frac{d \Omega_{\bbox{k}} d \Omega_{\bbox{q}}}{(4\pi)^{2}}
n^{3}_{\alpha}(\hat{\bbox{k}}) 
[B_{\alpha \beta}(\hat{\bbox{k}},\hat{\bbox{q}},0) -
 B_{\alpha \beta}(\hat{\bbox{k}},\hat{\bbox{q}},t) ] 
n^{3}_{\beta}(\hat{\bbox{q}})
\enspace.
\end{equation}
\\
Further terms in Eq.~(\ref{2.22}) containing coefficients $W^{\alpha}_{lm}$
are small, due to the assumption that 
$ B_{\alpha \beta}(\hat{\bbox{k}},\hat{\bbox{q}},0) \gg 
B_{\alpha \beta}(\hat{\bbox{k}},\hat{\bbox{q}},0) -
 B_{\alpha \beta}(\hat{\bbox{k}},\hat{\bbox{q}},t)$, and can be
 neglected. The same applies to Fick's law in 
Eq.~(\ref{2.10}), so that we arrive at the diffusion equation with
absorption which we already formulated in the introduction.

Using Eq.~(\ref{2.20}) for director modes we calculate

\begin{equation}
\mu(t) = \mu_{0} t \quad \mathrm{with} \quad \mu_{0} =
\frac{2k_{B}T}{9\pi}\, \frac{\omega^{4}}{c^{3}} \, 
\frac{(\Delta \varepsilon)^{2}}{\sqrt{\varepsilon_{\perp}}} \,
\frac{\widetilde{\mu}}{\gamma} \enspace,
\end{equation}
\\
where the numerical factor $\widetilde{\mu}$ stands for a dimensionless
angular average over the geometrical factor $N(\alpha,\beta,\delta)$ and 
the viscosities $\eta_{\delta}(\bbox{q}_{s}) / \gamma$. The factor
$\widetilde{\mu}$ is of the order of 1 in thermotropic nematics
but always larger than 1 
[20].
%\cite{Stark96b}.
Note that $\mu(t)$ only depends on viscosities and not at all on
elastic properties. They cancel because they determine both static
light scattering and hydrodynamics of the director modes.
For 5CB, $\gamma = 0.81\,\mathrm{P}$ which agrees very well with
the experimentally determined value of
$\gamma / \widetilde{\mu} = 0.60 \pm 0.20\,\mathrm{P}$ 
[21,\ 25].
%\cite{Stark97,Kao96}.
In experiments one fits the autocorrelation function with the help of
a numerical or exact solution of the diffusion equation under
the appropriate boundary conditions 
[7].
%\cite{Weitz92}.

Down to the experimental resolution of $4 \times 10^{-8} \, \mathrm{s}$
no deviation of the director dynamics from the Leslie-Erickson theory
was observed. It should show up in a different temporal power law
of $\mu(t)$. It would be interesting to study systems with higher
viscosities like polymer liquid crystals and to look for such a
deviation. The Brownian motion of colloidal particles in a fluid e.g. 
clearly does not show a simple diffusive behavior on very short time 
scales 
[7,\ 37].
%\cite{Weitz92,Kao93}.

%\section*{\normalsize \textbf{4. CONCLUSION}}

\subsection*{\normalsize \textbf{Acknowledgements}}
I thank Bart van Tiggelen, Roger Maynard, Georg Maret, Michael
Heckmeier and Martin \v Copi\v c for stimulating discussions.
Tom Lubensky, Arjun Yodh, Ming Kao and Kristen Jester contributed
considerably to develop my understanding of diffusing light in nematics.

\subsection*{\normalsize \textbf{References}}
%\begin{thebibliography}{10}

\chead{H. STARK}

\begin{tabbing}
%\bibitem{Kuga84}
[10]\=   \kill
\enspace[1]\> \enspace 
Y. Kuga and A. Ishimaru, {\em J.~Opt.\ Soc.\ Am.~A}, {\bf 1},  831  (1984);
M.~P.\linebreak\\
 \> \enspace van Albada and A. Lagendijk, {\em Phys.\ Rev.\ Lett.}, {\bf
  55},  2692  (1985);\\
 \> \enspace 
P.-E. Wolf and G. Maret, {\em Phys.\ Rev.\ Lett.}, {\bf 55},  2696  (1985).\\
%\bibitem{Sheng90}
\enspace[2]\> \enspace
{\em Scattering and Localization of Classical Waves in Random Media},\\
 \> \enspace
edited by P. Sheng (World Scientific, Singapore, 1990).\\
%\bibitem{Anderson58}
\enspace[3]\> \enspace
P.~W. Anderson, {\em Phys. Rev.}, {\bf 109},  1492  (1958).\\
%\bibitem{Golubentsev84}
\enspace[4]\> \enspace
A.~A. Golubentsev, {\em Sov.\ Phys.\ JETP}, {\bf 86},  26  (1984);
E. Akkermans\\
 \> \enspace
and R. Maynard, {\em J.~Physique Lett.}, {\bf 46},  L-1045 (1985);
E. Akker-\\
 \> \enspace
mans, P.~E. Wolf, and R. Maynard, {\em Phys.\ Rev.\ Lett.}, {\bf 56},  1471
 (1986).\\
%\bibitem{Maret87}
\enspace[5]\> \enspace
G. Maret and P.~E. Wolf, {\em Z.\ Phys.~B}, {\bf 65},  409  (1987).\\
%\bibitem{Pine88}
\enspace[6]\> \enspace
D.~J. Pine, D.~A. Weitz, P.~M. Chaikin, and E. Herbolzheimer, 
{\em Phys.}\\
 \> \enspace
{\em Rev.\  Lett.}, {\bf 60},  1134  (1988).\\
%\bibitem{Weitz92}
\enspace[7]\> \enspace
D.~A. Weitz and D.~J. Pine,  in {\em Dynamic Light Scattering}, edited
by\\
 \> \enspace
 W. Brown (Oxford University Press, Oxford, 1992), pp.\ 652--720.\\
%\bibitem{Wu90}
\enspace[8]\> \enspace
X.~L. Wu {\it et~al.}, {\em J. Opt. Soc. Am. B}, {\bf 7},  15  (1990).\\
%\bibitem{Gang94}
\enspace[9]\> \enspace
H. Gang, A.~H. Krall, and D.~A. Weitz, {\em Phys. Rev. Lett.}, {\bf
  73},  3435\\
 \> \enspace
  (1994).\\ \protect
%\bibitem{Durian91}
[10]\> \enspace
D.~J. Durian, D.~A. Weitz, and D.~J. Pine, {\em Science}, {\bf 252},  617
(1994).\\ \protect
%\bibitem{Boas95}
[11]\> \enspace
D.~A. Boas, L.~J. Campbell, and A.~G. Yodh, {\em Phys. Rev. Lett.},
{\bf75},\\
 \> \enspace
 1855 (1995).\\ \protect
%\bibitem{Heckmeier96}
[12]\> \enspace
M. Heckmeier and G. Maret, {\em
  Europhys. Lett.}, {\bf 34}, 257 (1996); M.\\
 \> \enspace
Heckmeier, S.~E.\ Skipetrov, G.\ Maret, and R.\ Maynard, {\em
  J. Opt. Soc.}\\
 \> \enspace
 {\em Am. A}, {\bf 14}, 185 (1997).\\ \protect
%\bibitem{Yodh95}
[13]\> \enspace
A.~G. Yodh and B. Chance, {\em Physics Today}, {\bf 48},  34
(1995).\\ \protect
%\bibitem{Tiggelen95}
[14]\> \enspace
B.~A.\ van Tiggelen, {\em Phys. Rev. Lett.}, {\bf 75}, 422 (1995);
G.~L.~J.~A.\ Rik-\\
 \> \enspace
 ken and B.~A.\ van Tiggelen, {\em Nature (London)}, {\bf 381}, 54
 (1996).\\ \protect
%\bibitem{Vlasov88}
[15]\> \enspace
 D.~V. Vlasov, L.~A. Zubkov, N.~V. Orekhova, and V.~P. Romanov,\\
 \> \enspace
 {\em JETP\ Lett.}, {\bf 48}, 91 (1988).\\ \protect
%\bibitem{Vithana93}
[16]\> \enspace
H.~K.~M. Vithana, L. Asfaw, and D.~L. Johnson, {\em Phys.\ Rev.\ Lett.},\\
 \> \enspace
 {\bf 70}, 3561 (1993).\\ %\protect
%\bibitem{Gennes93}
\chead{LIGHT DIFFUSION IN NEMATICS}
$\!\!$[17]\> \enspace
P.~G. de\ Gennes and J. Prost, 2nd  ed. 
{\em The Physics of Liquid Crys-}\\
 \> \enspace
 {\em tals} (Clarendon  Press, Oxford, 1993).\\ \protect
%\bibitem{Chandrasekhar92}
[18]\> \enspace
S. Chandrasekhar, {\em Liquid Crystals}, 2nd  ed. (Cambridge University\\
 \> \enspace
Press, Cambridge, 1992).\\ \protect
%\bibitem{Stark96a}
[19]\> \enspace
H. Stark and T.~C. Lubensky, {\em Phys.\ Rev.\ Lett.}, {\bf 77}, 2229
(1996).\\ \protect
%\bibitem{Stark96b}
[20]\> \enspace
 H. Stark and T.~C. Lubensky, {\em Phys.\ Rev.\ E}, {\bf 55},
514 (1997).\\ \protect
%\bibitem{Stark97}
[21]\> \enspace
H. Stark, M.~H. Kao, K.~A. Jester, T.~C. Lubensky, A.~G. Yodh, and\\
 \> \enspace
P.~J. Collings, {\em J. Opt. Soc. Am. A}, {\bf 14}, 156 (1997).\\ \protect
%\bibitem{Tiggelen96}
[22]\> \enspace
B.~A.\ van Tiggelen, R. Maynard, and A. Heiderich, {\em  Phys. Rev. Lett.},\\ 
 \> \enspace
{\bf 77}, 639 (1996); B.~A.\ van Tiggelen, A. Heiderich, and R. Maynard,\\
 \> \enspace
{\em Mol. Cryst. Liq. Cryst.}, 205 (1997).\\ \protect
%\bibitem{Heiderich97}
[23]\> \enspace
A. Heiderich, R. Maynard, and B.~A.\ van Tiggelen, {\em
  J. Phys. (France)}\\
 \> \enspace
 {\em II}, {\bf 7}, 765 (1997).\\ \protect
%\bibitem{Tiggelen98}
[24]\> \enspace
B.~A. van Tiggelen and H. Stark, {\em Int. J. Mod. Phys. B} 
(to be pub-\\
 \> \enspace
lished).\\ \protect
%\bibitem{Kao96}
[25]\> \enspace
M.~H. Kao, K. Jester, A.~G. Yodh, and P.~J. Collings, {\em
  Phys. Rev.}\\
 \> \enspace
{\em Lett.}, {\bf 77}, 2233 (1996).\\ \protect
%\bibitem{Schuster05}
[26]\> \enspace
A. Schuster, {\em Astrophys.~J.}, {\bf 21},  1  (1905).\\ \protect
%\bibitem{Ishimaru78}
[27]\> \enspace
A. Ishimaru, {\em Wave Propagation in Random Media}, Vols. 1 and 2,\\
 \> \enspace
 (Academic, New York, 1978).\\ \protect
%\bibitem{Hulst80}
[28]\> \enspace
H.~C. van de Hulst, {\em Multiple Light Scattering}, Vols. 1 and 2,
(Acade-\\
 \> \enspace
mic, New York, 1980).\\ \protect
%\bibitem{Romanov88}
[29]\> \enspace
V.~P.\ Romanov and A.~N.\ Shalaginov, {\em Opt. Spectrosc. (USSR)},
{\bf 64}, \\
 \> \enspace
 774 (1988).\\ \protect
%\bibitem{Bellini96}
[30]\> \enspace
T. Bellini and N.~A. Clark, in {\em Liquid Crystals in Complex
  Geome-}\\
 \> \enspace
{\em tries}, edited by G.~P.\ Crawford and S.\ \v Zumer (Tayler \&
Francis, Lon-\\
 \> \enspace
don, 1996), Chap. 19, p. 381.\\ \protect
%\bibitem{Drzaic95}
[31]\> \enspace
P.~S. Drzaic, {\em Liquid Crystal Dispersions} (World Scientific,
Singapore,\\
 \> \enspace
 1995).\\ \protect
%\bibitem{Yang96}
[32]\> \enspace
D.-K. Yang, L.-C. Chien, and Y.~K. Fung, in {\em Liquid Crystals in
  Com-}\\
 \> \enspace
 {\em plex
  Geometries}, edited by G.~P.\ Crawford and S.\ \v Zumer (Tayler \&\\
 \> \enspace
Francis, London, 1996), Chap. 5, p. 103.\\ \protect
%\bibitem{Landau60}
[33]\> \enspace
L.~D. Landau and E.~M. Lifschitz, {\em Electrodynamics of Continuous}\\
 \> \enspace
 {\em Media},
  Vol.~8 of {\em Course of Theoretical Physics}, first englisch ed. \\
 \> \enspace
 (Pergamon Press, Oxford, 1960).\\ \protect
%\bibitem{Lax71}
[34]\> \enspace
M. Lax and D.~F. Nelson, {\em Phys.\ Rev.~B}, {\bf 4},  3694
(1971).\\ \protect
%\bibitem{Langevin75}
[35]\> \enspace
D. Langevin and M.-A. Bouchiat, {\em J.~Phys. (Paris) Colloq.}, {\bf 36},  
C1-\\
 \> \enspace
  197 (1975); A.~Y. Val'kov and V.~P. Romanov, {\em Sov.\ Phys.\ JETP}
  {\bf 63},\\
 \> \enspace  
737 (1986).\\ \protect
%\bibitem{Ackerson92}
[36]\> \enspace
Dougherty {\em et al.}, {\em J.\ Quant. Spectrosc. Radiat. Transfer},
{\bf 52}, 713\\
 \> \enspace
 (1994).\\ \protect
%\bibitem{Kao93}
[37]\> \enspace
M.~H. Kao, A.~G. Yodh, and D.~J. Pine, {\em Phys.\ Rev.\ Lett.}, {\bf 70},  
242\\
 \> \enspace
 (1993).
\end{tabbing}
\chead{H. STARK}
%\end{thebibliography}
\end{document}